**Title: Evolving Horizons in Radiotherapy Auto-Contouring: Distilling Insights, Embracing Data-Centric Frameworks, and Moving Beyond Geometric Quantification**

Authors: Kareem A. Wahid, PhD[1,2], Carlos E. Cardenas, PhD[3], Barbara Marquez[4,5], Tucker J. Netherton PhD, DMP[5], Benjamin H. Kann, MD[6], Laurence E. Court, PhD[5], Renjie He, PhD[2], Mohamed A. Naser PhD[2], Amy C. Moreno, MD[2], Clifton D. Fuller, MD, PhD[2], David Fuentes, PhD[1]*

[1] Department of Imaging Physics, The University of Texas MD Anderson Cancer Center, Houston, TX, USA.

[2] Department of Radiation Oncology, The University of Texas MD Anderson Cancer Center, Houston, TX, USA.

[3] Department of Radiation Oncology, University of Alabama at Birmingham, Birmingham, AL, USA.

[4] UT MD Anderson Cancer Center UTHealth Houston Graduate School of Biomedical Sciences, Houston, TX, USA.

[5] Department of Radiation Physics, University of Texas MD Anderson Cancer Center, Houston, TX, USA.

[6] Department of Radiation Oncology, Brigham and Women's Hospital, Dana-Farber Cancer Institute, Harvard Medical School, Boston, MA, USA.

* corresponding author, contact info: dtfuentes@mdanderson.org.

**Funding Statement:** KAW was supported by an Image Guided Cancer Therapy (IGCT) T32 Training Program Fellowship from T32CA261856. BM was supported by the American Legion Auxiliary Fellowships in Cancer Research and the UTHealth Innovation for Cancer Prevention Research Training Program Pre-doctoral Fellowship (Cancer Prevention and Research Institute of Texas grant #RP210042). BHK receives funding from NIH/NIDCR K08 DE030216. LEC has received funding from Varian Medical Systems, Wellcome Trust, Fund for Innovation in Cancer Informatics, and The Cancer Prevention and Research Institute of Texas. ACM receives unrelated funding and salary support from NIH National Institute of Dental and Craniofacial Research Exploratory/Developmental Research Grant Program (R21DE031082-01) and Mentored Career Development Award to Promote Diversity (K01DE030524-01A1), and infrastructure support from MD Anderson Cancer Center via the Charles and Daneen Stiefel Center for Head and Neck Cancer Oropharyngeal Cancer Research Program. CDF was supported by P30CA016672. DF was supported by R01CA195524.

**Conflicts of Interest:** CDF has received travel, speaker honoraria and/or registration fee waivers unrelated to this project from: The American Association for Physicists in Medicine; the University of Alabama-Birmingham; The American Society for Clinical Oncology; The Royal Australian and New Zealand College of Radiologists; The American Society for Radiation Oncology; The Radiological Society of North America; and The European Society for Radiation Oncology. The other authors have no interests to disclose.

**Introduction**

Historically, clinician-derived contouring of tumors and healthy tissues has been crucial for radiotherapy (RT) planning. In recent years, advances in artificial intelligence (AI), predominantly in deep learning (DL), have rapidly improved automated contouring for RT applications, particularly for routine organs-at-risk [1–3]. Despite research efforts actively promoting its broader acceptance, clinical adoption of auto-contouring is not yet standard practice.

Notably, within several AI communities, there has been growing enthusiasm to shift from conventional "model-centric" AI approaches (i.e., improving a model while keeping the data fixed), to "data-centric" AI approaches (i.e., improving the data while keeping a model fixed) [4]. Although balancing both approaches is typically ideal for crafting the optimal solution for specific use cases, most research in RT auto-contouring has prioritized algorithmic modifications aimed at enhancing quantitative contouring performance based on geometric (i.e., structural overlap) indices [5] — a clear testament to the "model-centric" AI paradigm.

In this editorial, aimed at clinician end-users and multidisciplinary research teams, we harmonize key insights in contemporary RT auto-contouring algorithmic development to motivate the adoption of data-centric AI frameworks for impactful future research directions that would further facilitate clinical adoption. Of note, the discussion herein draws primarily from literature related to head and neck cancer (HNC), showcasing it as a representative example of a complex disease site. However, these insights apply broadly to auto-contouring across disease sites.

**Insight 1: DL auto-contouring algorithms require high-quality training data**

The adage "garbage in, garbage out" is often used to describe the importance of providing computational algorithms with high-quality data (i.e., "ground truth"). One particular challenge for RT contouring applications is the absence of a definitive ground truth. In contouring research, ground truth typically refers to a structure delineated by a clinician, preferably with expertise in the relevant disease site. Ideally, this structure should show minimal differences if another expert were to contour it independently (i.e., low interobserver variability), given the observers desire the same clinical endpoint. Despite increasing guideline recommendations over time [6], some structures, such as target volumes, are inherently more subjective than others due to clinical factors and institutional preferences. Notably, the precise definition of ground truth in contouring is debated, as multiple clinically acceptable solutions for a single structure may exist [5,7]. Building on this context, a tangible manifestation of the "garbage in, garbage out" principle within HNC contouring is exemplified in a study by Henderson et al. [8]. Their findings revealed that models trained on a small set of consistent contours (i.e., strictly following guidelines) aligned more closely with the ground truth test data than those trained on a vast array of inconsistent contours (**Figure 1**). This underscores the critical role of consistent, high-quality contours for successful DL auto-contouring training.

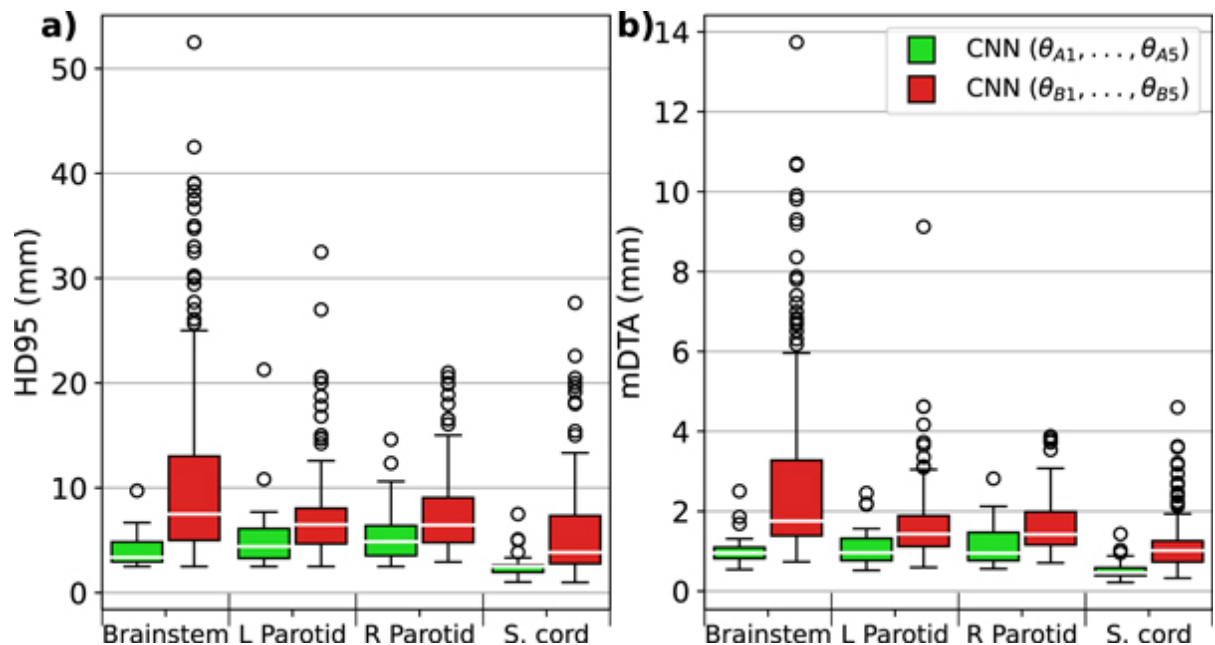

**Figure 1.** A deep learning model trained with a few highly consistent, i.e., high-quality, contours (green) was more closely aligned to the ground truth test data than a model trained with many inconsistent contours (red) for various head and neck cancer radiotherapy structures. The 95% Hausdorff distance (HD95) (**a**) and mean distance to agreement (mDTA) (**b**) were used as geometric performance quantification metrics. Lower values for both metrics indicate better performance. Reprinted from Henderson et al. [8].

Curating high-quality ground-truth contouring data is costly in terms of dedicated clinician effort. Expert clinicians must meticulously manually contour structures and, when applicable, carefully consider existing guidelines to reduce interobserver variability. Consensus contouring fusion methods, such as the Simultaneous Truth and Performance Level Estimation algorithm, have allowed for potentially suboptimal contours (e.g., deviating from guidelines) to be combined to yield an improved overall contour structure. Recent work by Lin et al. [9] investigated consensus methods across various RT disease sites using an unprecedented number of physician observers and revealed that as few as two to five non-expert contours can approximate expert gold standard geometric benchmarks (**Figure 2**). Conceivably, these consensus inputs could be cost-effective alternatives to expert-derived ground truth for DL auto-contouring training. In other words, institutions without access to established experts may still be able to produce high-quality data for algorithmic development.

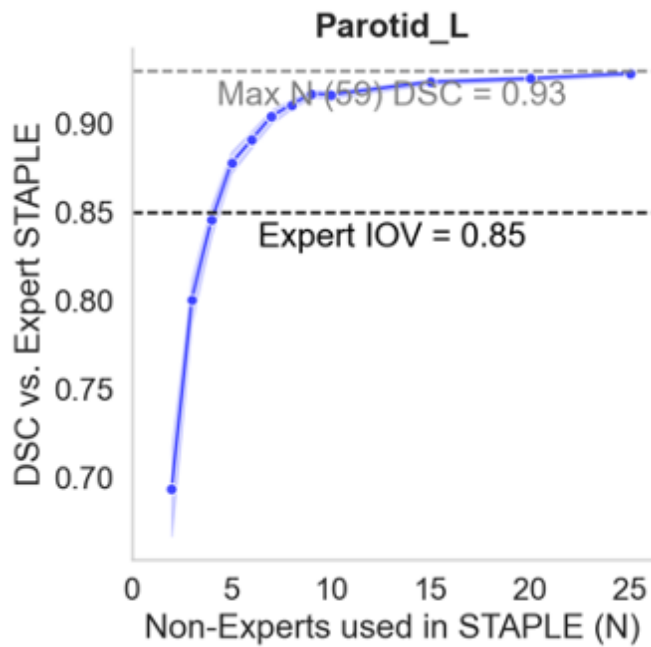

**Figure 2.** Consensus from a limited number of non-expert contours can approximate expert benchmarks. Specific plot is shown for the left parotid gland in a head and neck cancer case using the volumetric Dice similarity coefficient (DSC) as a performance quantification metric. The Simultaneous Truth and Performance Level Estimation (STAPLE) algorithm was used to generate consensus contours. To explore consensus quality dynamics based on the number of non-expert inputs, bootstrap resampling selected random non-expert subsets with replacement to form consensus contours, which were then compared to expert consensus. Each dot represents the median from 100 bootstrap iterations with a 95% confidence interval (shaded area). The black dotted line indicates the median expert DSC interobserver variability (IOV). The gray dotted line indicates DSC performance for the maximum number of non-experts used in the consensus. For this example, three to four non-experts can approximate expert IOV benchmarks. As the number of non-experts in the consensus contour increases, performance generally improves before plateauing. Adapted from Lin et al. [9].

**Insight 2: DL auto-contouring models exhibit reasonable quantitative performance with limited data**

While natural images (e.g., photographs) are abundant and simple to annotate, medical image contouring data are significantly limited. This has constrained DL contouring research in the medical image domain to much smaller training set sizes compared to their natural image counterparts. Nonetheless, DL auto-contouring models seemingly perform quite well in terms of geometric indices despite limited medical image training set sizes, assuming high-quality data. A study by Fang et al. [10] highlighted this phenomenon by showing most HNC organs-at-risk reach 95% of their maximum possible geometric performance with as few as 40 independent patient samples (**Figure 3**). Depending on context-specific use cases for certain structures, the appropriate sample sizes may be even smaller. Moreover, the study illustrated diminishing returns in quantitative performance with increasing training set size, noting that performance plateaus or even declines in some instances. Similarly, Yu et al. [11] and Weissmann et al. [12] demonstrate that small, well-curated datasets can be used to train publicly available models to achieve clinically acceptable results.

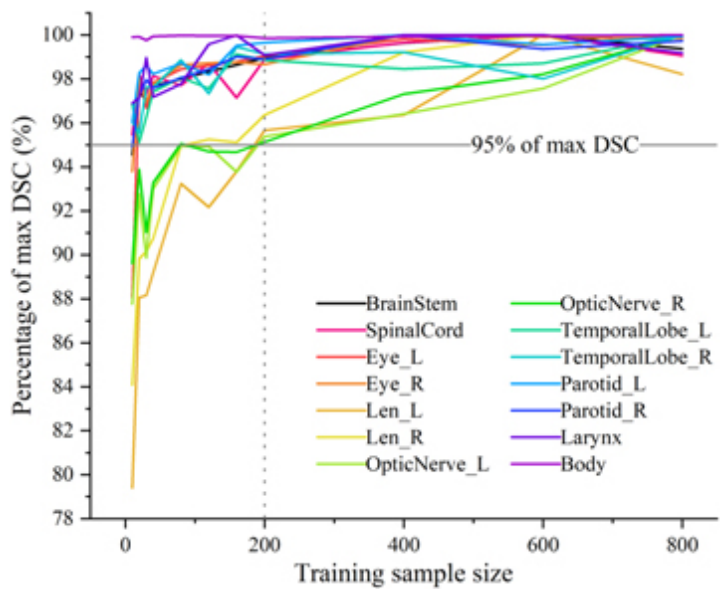

**Figure 3.** Relatively small training sample sizes are needed to reach high geometric performance for deep learning auto-contouring models. The percentage of the volumetric Dice similarity coefficient (DSC) using different training sample sizes relative to the maximum DSC for individual contour structures is shown in different colors. Most organ-at-risk structures required ~40 patient samples to achieve 95% of the maximum possible performance; notably, lenses and optic nerves required 200 samples to achieve 95% of the maximum possible performance. Reprinted from Fang et al. [10].

While DL models were historically labeled as "data hungry", modern approaches now allow them to perform impressively well even with what might appear as limited data. In auto-contouring, because training is fundamentally conducted at the scale of voxels, even modest patient populations can provide sufficient datasets for pattern learning. Notably, data-centric pre-processing strategies, such as performing image cropping to minimize the imbalance between "positive" and "negative" voxels before model training, further enhance this ability in auto-contouring [13].

**Insight 3: Auto-contouring quantitative performance is saturating**

The democratization of science, particularly through open-source tools and data, has justifiably become more prevalent over time. Much of this shift has also influenced the realm of radiotherapy research [14] and, by extension, medical image contouring. This has allowed for an increasingly "level" playing field for researchers in terms of algorithmic development. Within contouring, a prime example of the benefits of open-science practices has been the increasing use of U-Net, an effective DL contouring architecture, through standard computational libraries. nnU-Net [15], a self-configuring variant of the U-Net architecture, has unmistakenly become a de facto standard for many medical image contouring projects. More recently, the publicly available Segment Anything Model, which has been benchmarked on medical imaging data [16], has also yielded impressive results with minimal domain-specific training.

Over the past several years, medical image data challenges (i.e., public competitions), have been inundated with U-Net variants [15]. This surge has seemingly decreased the gap between 'state-of-the-art' and 'average' participant performance. In RT contouring, the HECKTOR challenge [17] — a competition focused on HNC gross tumor volume contouring using PET/CT imaging — stands out as a prime example, where the state-of-the-art contouring performance has steadily plateaued after median performance crossed expert interobserver variability (**Figure 4**). Moreover, once a measure of human performance benchmarking has been exceeded (e.g., interobserver variability), the practical benefits of further improving geometric indices become somewhat ambiguous. For particularly noisy contouring targets like tumor volumes, where human agreement on what constitutes an "acceptable" contour would already be low, the value of greater geometric performance optimization merits reconsideration.

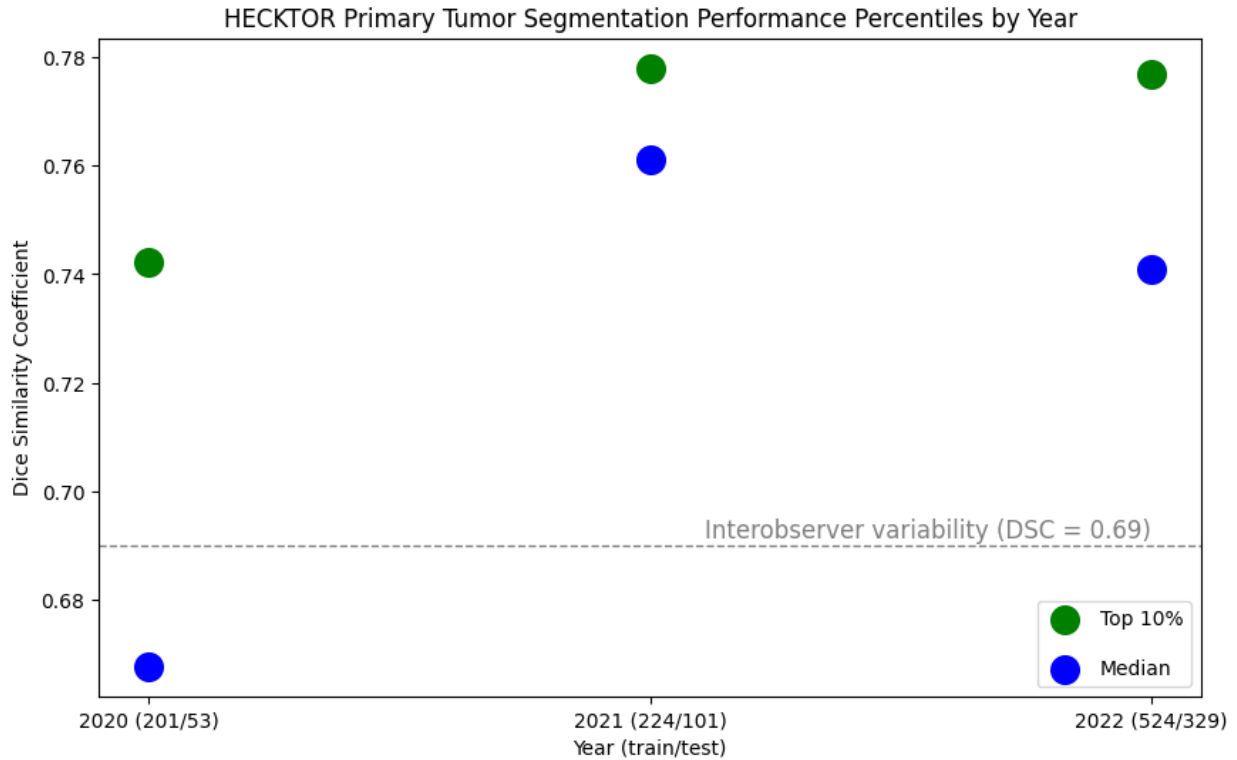

**Figure 4.** HEad and neCK TumOR (HECKTOR) contouring performance saturation. Contouring performance measured by volumetric Dice similarity coefficient. Green and blue dots correspond to the top 10% and median tumor contouring performance measured across all participating teams, respectively. The gray dotted line corresponds to a clinician expert interobserver variability benchmark. Data derived from corresponding HECKTOR conference proceedings.

**Future perspectives on auto-contouring**

From the previous discussion points, it becomes increasingly clear that DL auto-contouring requires data that, perhaps contrary to popular belief, is surprisingly simple to curate. Moreover, given the open-source nature of state-of-the-art DL architectures, training these models is also seemingly straightforward. One could ostensibly collect a relatively small group of non-expert contours and generate consensus data to train a nnU-Net model that delivers reasonable geometric performance. So, is RT auto-contouring effectively a solved problem at this point? Though some facets of contemporary research seem to support this idea, there remain significant avenues of exploration before we can confidently say yes.

Most auto-contouring research has focused on geometric indices (e.g., volumetric Dice) as evaluation criteria [5], likely because these indices are commonly embedded within model training schemes. While geometric indices can serve as intuitive adjuncts for roughly gauging clinical acceptability, they are not a panacea. Geometric indices have been found to not be strongly correlated to dosimetric or clinical endpoints [7,18,19], so their utility in RT is potentially limited. A growing number of studies have started incorporating clinician-derived qualitative scoring evaluations, which may be more closely linked to clinical usability, but these methods may be prone to human bias [7]. Nonetheless, model-centric AI approaches that seek to gain increasingly diminishing returns in geometric performance by simply tweaking underlying DL architectures may not offer significant clinical benefits. Of note, it is not this editorial's intention to dissuade researchers from continuing investments in model-centric approaches but rather to emphasize the importance of assessing whether such endeavors lead to meaningful clinical impact. For example, recent model-centric approaches have demonstrated state-of-the-art contouring performance can be achieved by intelligently reducing the number of model parameters [20], thereby accelerating training and facilitating deployment in resource-constrained settings. Moreover, for challenging tumor-related structures there might still be room for improvement in geometric performance. However, one must question: would an improvement in a Dice score of 1% for say, a parotid gland contour, offer any tangible benefit? The clinical influence of such a change is doubtful. Future research is likely to explore alternative indices for quantification, particularly those that can accurately capture dosimetric impact.

Given the widespread availability of standardized auto-contouring DL architectures, there is a natural inclination for auto-contouring research to transition toward data-centric approaches. Additionally, unlike other industries where vast data repositories exist, medical research is marked by a relative data shortage [21], making the pursuit of a data-optimization strategy potentially more fruitful than model-optimization in the current landscape. For instance, fields of data-centric AI, like active learning, where models iteratively learn through user interaction, could be used to improve performance and minimize contouring time. Notably, interactive contouring has already been shown to be clinically feasible for HNC tumors [22] and OARs [23]. Furthermore, as additional imaging modalities like magnetic resonance imaging become relevant for RT planning [24], data-centric AI methods such as domain adaptation and transfer learning — techniques that apply knowledge from one data environment to another — are anticipated to rise in prominence. Illustrating these concepts, Boyd et al. [25] adapted a glioma auto-contouring model from an adult to a pediatric population, thereby demonstrating effective translation even in limited data scenarios. Moreover, data-centric techniques could, given appropriate regulatory approval, conceivably be employed in the future to better tailor solutions to specific institutions or user preferences. Recent work by Balagopal et al. [26] demonstrated that a pre-trained auto-contouring model could be tailored to particular practice styles with only a limited amount of new data. This challenges the traditional objective of ensuring generalized performance across institutions to emphasize usability for individual entities, highlighting potentially evolving priorities in DL auto-contouring.

Importantly, literature within AI-augmented decision-making highlights the need to design support tools that align with clinicians' intended use. Recent evidence has shown clinicians do not fully capitalize on the potential gains from image-based AI assistance, even when these models consistently outperform experts [27]. Additionally, the challenge of automation over-reliance is expected to pose problems when users interact with these systems [28]. This underscores the imperative of increasing research into model uncertainty estimation and explainability methods [29]. Model uncertainty and explainability will likely become an increasingly relevant facet for ensuring clinician trust and engagement when implementing RT auto-contouring tools. Techniques that align model uncertainty with human expectations using data-centric approaches are poised to gain significance. Furthermore, as we increasingly rely on these models, ensuring they remain unbiased, particularly toward underrepresented or marginalized communities, is paramount. The consequences of biased AI can range from inaccurate predictions to reinforcing systemic inequalities [30]. Thus, adopting specific data-centric strategies focused on assuring representation and consistent performance will not just be beneficial — but a moral imperative.

**Conclusion**

Model-centric AI has made great strides in RT auto-contouring. Nevertheless, given DL auto-contouring facile training characteristics, readily available state-of-the-art architectures, and a plateauing of geometric performance, it becomes imperative for the auto-contouring community to pivot their focus. Embracing data-centric techniques, such as active learning and transfer learning, and exploring alternative methods to capture clinical utility, such as dosimetric impact and model uncertainty, could chart the next frontier in auto-contouring and allow for more facile clinical adoption. This shift not only recognizes the evolving needs and challenges of clinicians but also holds the promise of driving more clinically relevant breakthroughs for patients.

*Data availability statement*: Tabular data and Python code used to create the HECKTOR performance saturation figure are available on GitHub (https://github.com/kwahid/autoseg_editorial_code/tree/main).

*Declaration of generative AI and AI-assisted technologies in the writing process:* During the preparation of this work, the authors used ChatGPT (GPT-4 architecture; ChatGPT September 25 Version) to improve the grammatical accuracy and semantic structure of portions of the text. After using this tool, the authors reviewed and edited the content as needed and take full responsibility for the content of the publication.

**References**

1. Cardenas, C. E., Yang, J., Anderson, B. M., Court, L. E. & Brock, K. B. Advances in Auto-Segmentation. *Semin. Radiat. Oncol.* **29,** 185–197 (2019).

2. Santoro, M., Strolin, S., Paolani, G., Della Gala, G., Bartoloni, A., Giacometti, C., Ammendolia, I., Morganti, A. G. & Strigari, L. Recent Applications of Artificial Intelligence in Radiotherapy: Where We Are and Beyond. *NATO Adv. Sci. Inst. Ser. E Appl. Sci.* **12,** 3223 (2022).

3. Naqa, I. E. in *Artificial Intelligence in Radiation Oncology and Biomedical Physics* 1–23 (CRC Press, 2023).

4. Hamid, O. H. From Model-Centric to Data-Centric AI: A Paradigm Shift or Rather a Complementary Approach? in *2022 8th International Conference on Information Technology Trends (ITT)* 196–199 (2022).

5. Mackay, K., Bernstein, D., Glocker, B., Kamnitsas, K. & Taylor, A. A Review of the Metrics Used to Assess Auto-Contouring Systems in Radiotherapy. *Clin. Oncol.* **35,** 354–369 (2023).

6. Lin, D., Lapen, K., Sherer, M. V., Kantor, J., Zhang, Z., Boyce, L. M., Bosch, W., Korenstein, D. & Gillespie, E. F. A Systematic Review of Contouring Guidelines in Radiation Oncology: Analysis of Frequency, Methodology, and Delivery of Consensus Recommendations. *Int. J. Radiat. Oncol. Biol. Phys.* **107,** 827–835 (2020).

7. Baroudi, H., Brock, K. K., Cao, W., Chen, X., Chung, C., Court, L. E., El Basha, M. D., Farhat, M., Gay, S., Gronberg, M. P., Gupta, A. C., Hernandez, S., Huang, K., Jaffray, D. A., Lim, R., Marquez, B., Nealon, K., Netherton, T. J., Nguyen, C. M., Reber, B., Rhee, D. J., Salazar, R. M., Shanker, M. D., Sjogreen, C., Woodland, M., Yang, J., Yu, C. & Zhao, Y. Automated Contouring and Planning in Radiation Therapy: What Is 'Clinically Acceptable'? *Diagnostics* **13,** 667 (2023).

8. Henderson, E. G. A., Vasquez Osorio, E. M., van Herk, M., Brouwer, C. L., Steenbakkers,
9